# Exact Time-Dependent Solution to the Three-Dimensional Euler- Helmholtz and Riemann-Hopf Equations for Vortex Flow of a Compressible Medium and one of the Millennium Prize Problems

**S.G. Chefranov**[1) ] **and A.S. Chefranov**[2)]

[1,2)] Obukhov Institute of Atmospheric Physics of the Russian Academy of Sciences, Moscow, Russia

schefranov@mail.ru[1)]  a.chef@bk.ru[2)]

## Summary

For the first time the exact solution in unbounded space is represented for the two and three-dimensional Euler-Helmholtz (EH) vortex equation in the case of a nonzero-divergence velocity field for an ideal compressible medium flows. This solution corresponds to the obtained in Euler variables two and three-dimensional Riemann-Hopf (RH) equation solution. A necessary and sufficient condition of the onset of a singularity in the evolution of enstrophy in finite time $t = t_0$ for this solution is formulated. The exact closed description of the evolution of enstrophy and all other one-, two-, and multi point moments of the vortex field is now possible. Besides, the problem of closure, that is the main problem of the turbulence theory, is not arising here at all. The smooth continuation of the obtained solution to the EH an RH equations for $t \geq t_0$, due to introduction of a fairly large homogeneous friction or an arbitrary small effective viscosity, is stated. A new analytic, smooth for all time, solution to the two and three-dimensional Navier-Stokes (NS) equation is obtained. This solution coincides with the above-mentioned smooth solution to the EH and RH equations, which takes into account the viscosity effect of a compressible medium and also a known linear relation between the fluctuations of pressure and the velocity field divergence in systems far from the equilibrium. This gives the positive answer to the generalization of the known Millennium Prize Problem (www.claymath.org ) on the case of the compressible Navier-Stokes equation. Indeed, earlier only a negative answer for the problem of the existence of smooth solutions for any finite time has been a priori considered for the compressible case.

## Introduction

1. The understanding of many natural and technological processes is related to existence of the basic and applied problems of turbulence which remain unsolved already during more than hundred years by virtue of the absence of analytic time-dependent smooth vortical solutions to the Navier-Stokes (NS) equation. The development of the statistical approach to its solution gave many interesting results but also led to a new and so far unsolvable problem of closure in description of various moments of the vortex field whose approximate solution has been proposed by Kolmogorov, Heisenberg and other [1].

In fact, so far only a few exact solutions are known in hydrodynamics. However, none of these solutions is time-depended and defined in unbounded space (or in space with periodic boundary conditions) [1 - 3]. There are only weak time-dependent solutions which describe, for example, dynamics and interaction of singular vortical objects in a two- and three-dimensional ideal incompressible medium [2,4,5].

For three-dimensional flows of an ideal medium there is a concept of the possibility of existence of time-dependent solutions to the Euler-Helmholtz (EH) equation only on a bounded time interval $0 \le t < t_0$ (see [1,2,5,6] and the literature cited herein). In the case of the incompressible medium the length of this time interval is determined only by the three-dimensional effect of extension of the vortex filaments which can lead to explosive unbounded growth of the enstrophy (integral of the vorticity square over the space) in a finite time $t_0$ [1,2, 5, 6]. On the other hand, the exact steady-state regimes of flows of a viscous incompressible medium in the form of the Burgers and Sullivan vortices are well known [2]. For them the effect of extension of the vortex filaments potentially dangerous for the development of singularity is exactly compensated by the viscosity effect. However, for these solutions there exists no convergent energy integral over the entire unbounded space.

2. The question of existence of smooth time-dependent nonzero-divergence and divergence-free solutions to the three-dimensional NS equation in unbounded space (or in space with periodic boundary conditions) and on an unbounded time interval remains open during almost two hundred years (starting from 1827-1845) [3, 7 - 11]. The importance of this problem is determined by not only purely mathematical but also practical interest in connection with the basic and applied problem of predictability of hydro-meteorological and other fields developed in using the methods of numerical solution of the NS equation [8, 9].

In this connection, in 2000 the problem of existence of a smooth time-dependent vortical solution to the three-dimensional NS equation on an unbounded time interval was stated by the Clay Mathematics Institute as one of seven basic Millennium Prize Problems [7]. However, in [7] it was proposed to consider the solution of this problem not for the complete NS equation [3] but only for an equation obtained from the latter under the assumption on the divergence-free velocity field for the incompressible fluid. Evidently, in the choice of such a formulation a priori it was assumed that for flows with nonzero divergence of the velocity field the complete Navier-Stokes equation cannot deliberately have smooth solutions on the indefinite time interval. In fact, in [11] it was written about this: "The Millennium Prize Problem relates to incompressible flows since it is well known that the behavior of compressible flows is abominable". In this connection the example for a shock wave, developed in a compressible medium when a body moves through it at a velocity higher than the sonic speed of the medium, is given [11]. However, this does not exclude the possibility of existence of smooth solutions with nonzero divergence of velocity to the complete NS equation due to the effect of viscosity.

3. Therefore, the formulation of the problem given in [7] can be well generalized to include the case of flows with nonzero divergence of velocity for the compressible medium. This is the subject of the present study.

In fact, a new analytic time-dependent vortex solution to the complete 3-D NS equation is obtained here on the basis of exact vortex solution for the 3-D Riemann-Hopf (RH) equation and its modifications which takes into account the effective viscosity or friction [12-16] (see also Appendix A).

This solution remains smooth on any arbitrary large time intervals precisely owing to finiteness of the viscosity forces. In this case the solution to the NS equation can be continued in the Sobolev space $H^q(R^3)$ for any $q \geq 1$ and $t \geq t_0$, where $t_0$ is minimum time of the development of a singularity (collapse) for the corresponding exact solution to the RH equation at zero viscosity. For the Sobolev space $H^q(R^3)$ the norm is defined in the form [17]:

$$\|\vec{u}\|_{H^q(R^3)} = \left( \sum_{\beta \leq q} \int d^3x (\nabla^\beta \vec{u})^2 \right)^{1/2} \qquad \text{(I.1)}$$

In [17] the theorem of existence of a solution to the 3-D Euler equation (that is NS equation with zero viscosity) local in time was formulated for the divergence-free ideal incompressible fluid flow. In accordance with this theorem, a smooth solution to the Euler equation exists if the initial velocity field $\vec{u}_0$ belongs to the Sobolev space $H^q(R^3)$ when $q \geq 3$ with $\|\vec{u}_0\|_{H^3} \leq N_0 = const$ for a certain $N_0 > 0$. Then there exists a time $t_*(N_0) > 0$, which depends only on $N_0$, such that for all $t \in [0, t_*]$ the Euler equation has a solution belonging to the class $\vec{u} \in C([0,t_*]; H^q) \cap C^1([0,t_*]; H^{q-1})$, where the norm is defined in (I.1). Then, for the exact solution to the Euler-Helmholtz (EH) and RH equations considered in the present study (see Appendix A and Appendix B), in the case of ideal compressible fluid flow it is possible to continue this solution for times $t_* \geq t_0$ only in the Sobolev space $H^0(R^3)$ and already it is impossible to continue this solution in the Sobolev space $H^1(R^3)$ at time $t_* \geq t_0$, when threshold value $q = 1$ instead of the $q = 3$ in [17].

The obtained analytic solution to the NS equation corresponds to a finite divergence of the velocity field. This indicates that a priori assumption on the absence of smooth solutions to the compressible 3-D NS equation is invalid.

Our way of taking the viscosity into account is a particular example of simulation of turbulence when a random velocity field is introduced instead of the random force [18]. However, in [18] only the large-scale random velocity field was considered and the drift part of this velocity dependent only on time was eliminated. Averaging over the random velocity dependent only on time just ensures the simulation of the effective viscosity in the present paper. In this case it is of importance that this way of simulation of the viscosity effect does not change the structure of the viscosity force $\vec{F}_\nu$ which enters into the NS equation and, for example, for the incompressible medium has the form $\vec{F}_\nu = \nu\Delta\vec{u}$ [3].

In fact, it is known [18] that the existence of a solution to the NS equation turns out to be proved if a term proportional to the higher derivative (of the flow velocity $\vec{u}$ ) of the form $\Delta^\alpha \vec{u}, \alpha \geq \frac{5}{4}$ (see [19, 20]) is added to the usual viscosity force $\vec{F}_\nu$. This term changes the structure of the viscosity force of the initial NS equation.

It is also shown that the elimination of singularity of the solutions to the EH and RH equations always takes place in introducing a fairly large homogeneous (or external) friction with coefficient $\mu$ which must satisfy the condition (5.3) and corresponds to the change $\nu\Delta\vec{u} \to -\mu\vec{u}$ in the NS equation.

The new solution to the three-dimensional NS equation is obtained providing the zero total balance of the normal stresses caused by the pressure and the viscosity of flow with nonzero divergence of velocity for a compressible medium. As shown in Section 2, this corresponds to the sufficient condition of positive definiteness of the integral entropy growth rate. This makes it possible to reduce solution of the NS equation to solution of a three-dimensional analog of the Burgers equation and then also to solution of the three-dimensional RH equation and its generalization to include the case of taking into account the viscosity forces (external friction and the above-mentioned effective friction related to the random velocity field).

We also note that in the general case the vortical solutions to the three-dimensional RH equation coincide with the solutions to the three-dimensional EH equation for describing vortex flows of an ideal compressible medium with nonzero divergence of the velocity field [9, 12-16] (see also Appendix A and Appendix B).

In fact, all real media are compressible and generally their flows must be described by precisely solutions to the compressible NS equation. On the other hand, for a conventionally incompressible medium the flows with nonzero divergence of velocity can correspond to the presence of distributed sources and stocks whose simulation is successfully used in non-relativistic and relativistic hydrodynamics [21-24].

4. We note that in [25] an exact solution to the three-dimensional RH equation, which describes the explosive time evolution of the matrix of the first derivatives of the velocity field only in the Lagrangian variables, was obtained. This does not make it possible to obtain exact solutions to the three-dimensional EH equation for the vortex field on its basis, as made in [12, 13] and here (see Appendix B) in the Euler representation of the exact solution of RH equation (see Appendix A). At the same time, in the present study it is shown that the exact solution to the 3-D RH equation for the velocity field obtained here (see formula (3.5) below and (A.11) in Appendix A) gives the expression (3.10) in the Lagrangian representation for the evolution of the matrix of the first derivatives of the velocity. This formula exactly coincides with a formula given in [25] (see formula (30) in [25]).

In the two- and three-dimensional cases new analytic solutions for the evolution of the intensities of vortex and of the helicity of the Lagrangian liquid particles are also obtained (see (4.4) – (4.6)). In [26] a structurally similar form (4.7) of the solution to the EH equation (see formula (23) in [26]) was considered on the basis of using a combination of the Eulerian and Lagrangian descriptions in representation of vortex lines. However, on the basis of theory [26] it does not possible explicitly to describe the time evolution of the vorticity and its higher moments, but this is possible due to our solution (4.5).

In the present study we obtained a new necessary and sufficient criterion of implementation of the explosive singularity (collapse) in finite time (see (3.7) and (3.8)) for the inviscid solution to the RH and EH equations in the one-, two-, and three-dimensional cases. At the same time, in [25] the integral criterion is given in the form (3.9) (see formula (38) in [25]). This criterion determines only the sufficient condition for implementation of collapse of the solution. For example, when the initial velocity field is divergence-free the collapse is possible only in accordance with the necessary and sufficient criterion (3.8) but already it cannot be established from the criterion of [25]. Moreover, from consideration of the explosive regime carried out in [25] for the solution to the three-dimensional RH equation a conclusion is made that this solution cannot be continued in infinite time in the Sobolev space $H^2(R^3)$. This differs from the above-mentioned result obtained in the present study.

In the two-dimensional case there is an exact correspondence between the criterion (3.7) and a similar criterion given in [27] (see formula (9) in [27]) in connection with the solution of the problem of propagation a flame front (generated by a self-sustained exothermal chemical reaction) on the basis of the simplified version of the Sivashisky equation [28]:

$$\frac{\partial f}{\partial t}-\frac{1}{2}U_s\left(\vec{\nabla}f\right)^2=\gamma_0 f \qquad \text{(I.2)}$$

In Eq. (I.2) the function $x_3=f(x_1,x_2,t)$ determines the flame front which represents the interface between a combustible matter ( $x_3>0$ ) and the combustion products ( $x_3<0$ ), $U_s,\gamma_0$ are constant positive quantities which characterize the front velocity and the combustion intensity, respectively. For $\gamma_0=0$ equation (I.2) coincides with the Hamilton-Jacobi equation for a free non-relativistic particle. In the two-dimensional case (more exactly, in its modification with account for the homogeneous friction with the coefficient $\mu$ when $\mu=-\gamma_0$), the exact solution (3.5) of the RH equation gives also the exact solution of Eq. (I.2). In this case the solution (3.5) describes a potential flow with velocity $\vec{u}=-U_s\vec{\nabla}f$ at zero vorticity.

5. An important result of the present study consists in obtaining a closed description of the time evolution of the enstrophy and any higher moments of the vortex field, as well as the velocity field in the 2-D and 3-D cases. This is reached on the basis of the corresponding analytic solutions to the EH, RH, and NS equations in the case of zero viscosity and in the case of taking the homogeneous friction and the effective viscosity into account. As a result, the problem of closure in the theory of turbulence is solved here exactly and not approximately, as usual [1].

In the present study this became possible only owing to a relatively simple dependence on the initial conditions for the exact solutions obtained for the EH and RH equations for the velocity field (3.5) and the vortex field ((4.1) and (4.2)). This is absent in the well-known exact solution to the Burgers equation obtained using the nonlinear Cole-Hopf transform.

In particular, owing to this achievement, on the basis of the exact solutions (4.2) of EH equation the following estimates for the integrals of the vorticity field can be obtained in the three-dimensional case in the neighborhood of the moment of singularity of the solution when $t\to t_0$:

$\Omega_{3(2m)} = \int d^3 x \vec{\omega}^{2m} \cong O(\frac{1}{(t_0 - t)^{2m-1}})$; $\Omega_{3(m)} = \int d^3 x \vec{\omega}^{m} \cong O(\frac{1}{(t_0 - t)^{m-1}})$, when $m = 1,2,3,...$ .

Hence there directly follows the inequality

$$\Omega_{3(2m)} / \Omega^2_{3(m)} \cong O\left(\frac{1}{t_0 - t}\right) >> 1; t \to t_0, \qquad (I.3)$$

which indicates strong intermittency of the vortex field in the neighborhood of singularity. We note that usually the inequality $\Omega_{3(2m)} >> \Omega^2_{3(m)}$ is assumed to be really fulfilled in the case of strong vortex intermittency [18]; however, earlier, it could not be obtained from the exact solution of the problem of closure in theory of turbulence, as made in obtaining the estimate (I.3).

6. In the final part of the present study the possibility of existence of not only the divergent but also smooth divergence-free solutions to the NS equation is discussed on the basis of an analysis of the exact closed solution to the enstrophy balance equation (5.6) and the variation rate of the integral kinetic energy in (6.1) - (6.4).

## 1. The Cauchy problem for the compressible Navier-Stokes (NS) equation

### 1.1 The reducing of the NS equation to modification of the Riemann-Hopf (RH) equation

In the general case the equations of motion of a compressible viscous medium can be represented in the following form [3]:

$$\frac{\partial u_i}{\partial t} + u_j \frac{\partial u_i}{\partial x_j} = \frac{\eta}{\rho} \Delta u_i - \frac{1}{\rho} \frac{\partial}{\partial x_i} (p - (\zeta + \frac{\eta}{3})(\frac{\partial u_k}{\partial x_k})); \Delta = \frac{\partial^2}{\partial x_k \partial x_k} \qquad (1.1)$$

$$\frac{\partial \rho}{\partial t} + \frac{\partial}{\partial x_k}(\rho u_k) = 0 \qquad (1.2)$$

From the form of the second term on the right-hand side of Eq. (1.1) it follows that for a viscous compressible flow the normal stresses are determined by not only the pressure but also by the divergence of the velocity field.

In (1.1) and (1.2) $u_i$ is the velocity of the medium, the summation convention is assumed to be applied to the recurring subscripts varying from 1 to $n$ (here, $n$ is the dimensionality of space and, in what follows, we will consider the cases of $n = 1,2,3$ ), and $p, \rho, \eta, \xi$ are the pressure, the density, and the constant viscosity and dilatational (second) viscosity coefficients, respectively [3].

The system of equations (1.1) and (1.2) having four equations for five unknowns is not closed. To get its closure, providing correctness of the Cauchy problem in the unbounded space, one more equation is needed. Usually, as such an equation, a state equation is used that relates the pressure and density [29, 30], or the pressure, density, and temperature (or entropy) [31]. In the latter case, the system (1.1), (1.2) is to be complemented with an equation for the temperature (or entropy) [31, 32]. However, for the non-equilibrium flow system with finite velocity, it looks problematic in general defining of the adequate form of any state equation without dependence of pressure on velocity or its derivatives.

Moreover, for relatively small flow velocities (compared to the speed of sound in the given medium), instead of the state equation, for the system of equations (1.1), (1.2) closure, the incompressible approximation is used when the velocity field is assumed divergent-free and pressure is strongly depended on velocity. This approximation is also reflected in the Millennium problem definition where only divergent-free NS equation solutions shall be considered [7].

Instead of these approximate closure methods for the system (1.1), (1.2), in the second section, from the sufficient condition of the positive-definiteness of the entropy increase rate, a linear relationship is derived between the pressure and the divergence of velocity for viscous compressible medium. Corresponding closing the system (1.1), (1.2), the fifth equation is as follows:

$$p = (\zeta + \frac{\eta}{3}) div\vec{u} \qquad (1.3)$$

Under (1.3), the second term in the right-hand side of the equation (1.1) turns to zero identically. This corresponds to the zero total balance of the normal stresses and in the limit $\varsigma >> \eta$ the expression (1.3) has a form typical for systems far from the equilibrium (see (81.4), (81.6) in [3]). The representation of this type for the pressure is also used in [3] instead of the medium equation of state.

In the result, the system (1.1)- (1.3) is already closed and for it the Cauchy problem can correctly defined in the unbounded space with any dimension number n.

Equation (1.1) under (1.3) becomes equivalent to the n-dimensional Burgers equation

$$\frac{\partial u_i}{\partial t} + u_l \frac{\partial u_i}{\partial x_l} = \frac{\eta}{\rho} \Delta u_i \qquad (1.4)$$

System of equations (1.2), (1.4) is already closed but rather complicated for being solved in quadrature or even in explicit analytical form for respective Cauchy problem. It is due to the presence of the dependence on the density in the right-hand side of (1.4) characterizing effect of the volumetric viscosity force.

Consequently, instead of the exact accounting for the viscosity forces in the equation (1.4), the approximate effect of the external friction force with constant coefficient $\mu \geq 0$ is used. This according with (1.3) allows getting the Cauchy problem solution for the 3-dimensional NS equation analytically on the base of exact vortex solution of 3-D Riemann-Hopf (RH) equation (see Appendix A), obtained from (1.4) when right-hand side of (1.4) is replaced by:

$$\frac{\eta}{\rho} \Delta u_i \to -\mu u_i, \qquad (1.5)$$

Another possibility, realized in Section 3, to obtain analytical solution of NS equation (1.1) when (1.3) takes place is the modeling of effective viscosity with help of introducing of the random Gaussian delta-correlated in time velocity field $\vec{V}(t)$ (by substitution $u_i \to u_i + V_i(t)$ in (1.4)) instead of right-hand side term in (1.4). It is important that the last representation used for the viscosity force does not lead to the introducing of the more rapid damping of the high-frequency harmonics (in the limit of large wave numbers) compared to that realized in the standard variant of the 3-dimensional NS equation (i.e. with the Laplacian in the first but not higher order). It matters for the possibility of backing of the inference of the existence at all times of smooth analytical solution for the 3-D compressible NS equation obtained herein.

Thus, if in (1.4) to make the substitution (1.5) then from (1.4), we shall get modification of the RH equation as follows

$$\frac{\partial u_i}{\partial t} + u_l \frac{\partial u_i}{\partial x_l} = -\mu u_i \qquad (1.6)$$

An expression for the external friction force in the right-hand side of (1.6) also corresponds to the approximate accounting for the viscosity force (described by the right-hand side of (1.4)) actually used in any numerical solution of the NS equation when with necessity cutting is made for large wave numbers or corresponding small scales $\lambda_0$ that corresponds to the substitution (1.5) where in (1.5) we have $\mu = \frac{\nu}{\lambda_0^2}, \nu = \frac{\eta}{\rho_0}, \rho_0 = \min \rho$.

In the three-dimensional case (*n*=3) after applying the curl operator to both sides of Eq. (1.6), we obtain the Euler–Helmholtz (EH) vortex equation in the modification with homogeneous friction force (for zero friction with $\mu = 0$ -see the equation EH (B.1) in Appendix B):

$$\frac{\partial \omega_i}{\partial t} + u_k \frac{\partial \omega_i}{\partial x_k} = \omega_k \frac{\partial u_i}{\partial x_k} - \omega_i div\bar{u} - \mu\omega_i. \qquad (1.7)$$

For zero dissipation equation EH (1.7) (when $\mu = 0$) can be also obtained direct from corresponding to (1.1) Euler equation in the case when the relation $rot(\frac{1}{\rho}\vec{\nabla}p) = 0$ is take place for non-zero divergence of velocity. Thus it is important to obtain the exact solution of (1.7) on the base of the exact solution of (1.6) (see Appendix A and Appendix B) because it gives analytical vortex solution of NS equation (1.1).

In particular, it is shown that for the case with zero external friction coefficient $\mu = 0$ the smooth exact solution of (1.6) and (1.7) obtained can exist only on a bounded time interval $0 \le t < t_0$ (the value of $t_0$ will be determined from the solution to Eq. (3.7)).

In what follows we will find a threshold value of the external friction coefficient (see (5.3)) so that for the greater value of the coefficient $\mu > \mu_{th} = \frac{1}{t_0}$ the solution of NS equation (1.1) (which is reduced to (1.6) or (1.7) on the base of (1.3) and (1.5)) becomes regular for any time. The more effective regularization for any time we obtain also in the case, when the random velocity field $V_i(t)$ with arbitrary small amplitude is introduced instead of external friction when $\mu = 0$.

**2. The statement of the Cauchy problem for the reduced NS system**

As in [7], but for the compressible medium we give here the next statement of the Cauchy problem for system (1.2) and (1.6) (or (1.7)), when equation (1.6) (or (1.7)) is obtained from NS equation (1.1) on the base of (1.3) and (1.5).

The equations (1.2), (1.6) and (1.7) describe the vortex motion of compressible medium in $R^n$ (n=1, 2 or 3). These equations are to be solved for an unknown velocity, vorticity $\omega_i(\vec{x},t) \in R^n; u_i(\vec{x},t) \in R^n$ and density $\rho(\vec{x},t) \in R$, defined for position $\vec{x} \in R^n$ and time $t \ge 0$. We pay attention here only to the compressible fluid filling all of $R^n$. In addition to equations (1.2) and (1.6), (1.7) we consider the initial conditions

$$\begin{aligned} &\vec{u}(\vec{x},0) = \vec{u}_0(\vec{x}); \rho(\vec{x},0) = \rho_0(\vec{x}) \\ &\vec{\omega}(\vec{x},0) = \vec{\omega}_0(\vec{x}) = rot\vec{u}_0 \end{aligned} \qquad (1.8)$$

Here $\vec{u}_0(x)$- is a given, $C^\infty$ vector field on $R^n$ and $\rho_0(\vec{x})$-is given $C^\infty$ scalar field on $R$.

For physically reasonable solutions of (1.2), (1.6) and (1.7) with conditions (1.8), we want to make sure $\vec{u}_0(\vec{x}), \rho_0(\vec{x})$ does not grow large as $|\vec{x}| \to \infty$. Hence we shall restrict attention to initial conditions that satisfy

$$\left|\frac{\partial^m \vec{u}_0(\vec{x})}{\partial \vec{x}^m}\right| or \left|\frac{\partial^m \rho(\vec{x})}{\partial \vec{x}_m}\right| < C_{mK}(1+|\vec{x}|)^{-K} \qquad (1.9)$$

on $R^n$ or $R$ -respectively for any non – negative m and K.

We accept a solution of (1.2) and (1.6) and (1.7) as physically reasonable only if it satisfies

$$\rho(\vec{x},t), \vec{u}(\vec{x},t), \vec{\omega}(\vec{x},t) \in C^{\infty}(R^n \times [0,\infty)) \qquad (1.10)$$

and bounded energy

$$\int_{R^n} d^n x |\vec{u}(\vec{x},t)|^2 \le C, \forall t \ge 0 \qquad (1.11)$$

The solution for the field of pressure $p$ is determined from (1.3).

In Section 3, we give an exact solution of the defined above Cauchy problem for the equation (1.6), and in Section 4 also an exact solution of the Cauchy problem for the equation (1.7) is obtained (for $\mu = 0$, and also for $\mu > 0$). In Section 5, a solution is given of the Cauchy problem for the equation (1.2) that follows from the solution of Cauchy problem for the equation (1.6) ( when $\mu = 0$ but accounting for the effective viscosity related with introduction of the random Gaussian delta-correlated in time velocity field $\vec{V}(t)$ ).

## 2. Energy and entropy balance equation

1. Usually, in consideration of the system of four equations (1.1), (1.2) for five unknown functions, a relation between the density and the pressure (equation of state of the medium) is appended in order to equalize the numbers of the equations and the unknown functions.

Instead of this method, we shall now derive a similar equation, closing the system (1.1), (1.2), for divergent flows of a compressible medium which will replace the condition of the divergence-free velocity field for incompressible fluid flows.

For this purpose we shall obtain the energy and entropy balance equations which follow both from (1.1), (1.2) and from usual thermodynamic relations [33]. In the case of a single-component medium these relations take the form [33] (see (14.3), (15.6), and (15.7) в [33]):

$$\varepsilon = Ts - \frac{p}{\rho} + \Phi \qquad (2.1)$$

$$-sdT + \frac{dp}{\rho} = d\Phi \qquad (2.2)$$

$$d\varepsilon = Tds + \frac{p}{\rho^2} d\rho \qquad (2.3)$$

In (2.1) - (2.3) $T$ is the temperature; $\varepsilon, s$, and $\Phi$ are the internal energy, the entropy, and thermodynamic potential or the Gibbs free energy (of unit mass of the medium), respectively [33]. The equation (2.3) directly follows from Eq. (14.3) in [29] and is in exact accordance with Eqs. (2.1) and (2.2) for any function $\Phi$. For the single-component medium considered, providing invariability of the particle number, we will assume that in (2.1) and (2.2) $d\Phi = 0$ or $\Phi = \Phi_0 = const$.

In what follows, we shall use Eq. (2.3) in the form (see also p. 272 in [3]):

$$\frac{\partial \varepsilon}{\partial t} = T\frac{\partial s}{\partial t} + \frac{p}{\rho^2}\frac{\partial \rho}{\partial t} \qquad (2.4)$$

2. From Eqs. (1.1), (1.2) we can obtain the following balance equation for the integral kinetic energy $E = \frac{1}{2}\int d^n x \rho u^2$ :

$$\frac{dE}{dt} = -\eta \int d^n x (\frac{\partial u_i}{\partial x_k})^2 + \int d^n x \left[ p - (\zeta + \frac{\eta}{3}) div\vec{u} \right] div\vec{u} \qquad (2.5)$$

For the divergence-free incompressible fluid flow the formula (2.5) coincides exactly with formula (16.3) in [3] and serves as a generalization of the latter to the case of compressible viscous fluid flow. To derive Eq. (2.5) it is sufficient to multiply (1.1) scalarly by the vector $\rho u_i$, multiply Eq. (1.2) by the scalar $\frac{\vec{u}^2}{2}$, add the expressions obtained, and integrate the result over the entire space.

We note that in the case of an ideal (inviscid) medium from (2.5) it follows that the integral kinetic energy is an invariant only for the divergence-free flows, while for the divergent flows the invariant is only the total integral energy $E_h = \int d^3x(\rho\frac{\vec{u}^2}{2} + \rho\varepsilon)$ assumed to be conserved also for the viscous medium [3].

3. On the basis of Eqs. (1.1), (1.2), (2.1) and (2.4) we shall now derive the total energy balance equation for a viscous compressible medium and the corresponding entropy balance equation. As distinct from the derivation given in [3], we shall directly use Eq. (2.1), written with allowance for the above-mentioned equality $\Phi = \Phi_0 = const$. As a result, taking (2.1) into account, from (2.4) we obtain:

$$\frac{\partial}{\partial t}(\rho\varepsilon) = T\frac{\partial}{\partial t}(\rho s) + \Phi_0\frac{\partial\rho}{\partial t} \qquad (2.6)$$

In Eq. (2.6) it is convenient to represent the second term on the right-hand side with allowance for (1.2) in the form $\Phi_0\frac{\partial\rho}{\partial t} = -div(\Phi_0\rho\vec{u})$.

In this case from (1.1), (1.2), and (2.6) we can obtain the following total energy balance equation:

$$\frac{\partial}{\partial t}(\rho\frac{\vec{u}^2}{2} + \rho\varepsilon) = -\frac{\partial}{\partial x_k}\left[u_k(\rho(\frac{\vec{u}^2}{2} + \Phi_0) + p - (\zeta + \frac{\eta}{3})div\vec{u}) - \eta\frac{\partial}{\partial x_k}(\frac{\vec{u}^2}{2})\right] + T(\frac{\partial}{\partial t}(\rho s) - \frac{B}{T}),$$

$$B = \eta(\frac{\partial u_i}{\partial x_k})^2 - \left[p - (\zeta + \frac{\eta}{3})div\vec{u}\right]div\vec{u} \qquad (2.7)$$

As in [3], taking into account the requirement of vanishing the derivative of the integral total energy with respect to time $\frac{d}{dt}E_h = 0$, we obtain the following entropy balance equation:

$$\frac{\partial}{\partial t}(\rho s) = \frac{B}{T}, \qquad (2.8)$$

where the expression $B$ is given in (2.7).

The energy and entropy balance equations (2.7) and (2.8) do not coincide with the equations given in [3] in formulas (49.3) and (49.4), respectively. However, from the balance equation (2.7) we can exactly obtain these equations (49.3) and (49.4), as well as the integral entropy balance equation (49.6) given in [3]. For this purpose, instead of Eq. (2.6), in (2.7) it is necessary to use the equivalent representation $\frac{\partial}{\partial t}(\rho\varepsilon) = (\varepsilon + \frac{p}{\rho})\frac{\partial\rho}{\partial t} + \rho T\frac{\partial s}{\partial t}$ (used in [3] regardless of (2.1) and the assumption on the equality $\Phi = \Phi_0 = const$). More essentially, in addition to above in order to obtain the coincidence between (2.7) and (49.7) in [3], it is necessary the pressure gradient in (2.7), in accordance with [3], to express in the form $\frac{\partial p}{\partial x_k} = \rho\frac{\partial}{\partial x_k}(\varepsilon + \frac{p}{\rho}) - \rho T\frac{\partial s}{\partial x_k}$ following from the thermodynamic relation (2.3) (by adding the term $\frac{dp}{\rho}$ to both sides of (2.3)). Such a thermodynamic representation for the pressure gradient entering into (2.7) (and into (1.1)) corresponds to the ordinary meaning of the pressure which completely describes the normal stresses for compressible and incompressible media only at zero viscosity. It does not correspond to that new meaning of the pressure which arises namely in the case of description of dynamics of a viscous compressible medium in (1.1) by virtue of appearance of additional normal stresses proportional to divergence of the velocity field (about this see also p. 275 in [3]).

In what follows, this statement on not quite adequate representation of the pressure gradient (in formulas (2.7) and (1.1)) on the basis of using the thermodynamic relation (2.3) will be confirmed by means of the fundamental relation (2.10), obtained below, between the time variation rate of the integral entropy and the integral kinetic energy. In fact, the relation (2.10) directly follows from (2.5) and the integral entropy balance equation written just in the form (2.9) on the basis of (2.8). On the other hand, this relation (2.10) cannot deliberately be obtained from (2.5) and the integral entropy balance equation in the form given in [3] (see (49.6) in [3]).

4. From the entropy balance equation (2.8) there follows the balance equation for the integral entropy $S = \int d^3 x \rho s$ in the form (for the sake of simplicity, as in (2.7) and (2.8), we omit the terms describing the fluxes caused by the temperature gradient):

$$\frac{d}{dt} S = \eta \int d^3 x \frac{1}{T} \left(\frac{\partial u_i}{\partial x_k}\right)^2 - \int d^3 x \frac{1}{T} div\vec{u} \left[ p - \left(\zeta + \frac{\eta}{3}\right) div\vec{u} \right] \tag{2.9}$$

As noted in the previous item, the balance equation (2.9) differs significantly from the integral entropy balance equation given in [3] (see formula (49.6) in [3]).

In the case of constant temperature $T = T_0$ in (2.9), from (2.9) and (2.5) there directly follows the exact fulfillment of the fundamental relation

$$T_0 \frac{dS}{dt} = -\frac{dE}{dt} \tag{2.10}$$

between the mechanical energy variation rate and the integral entropy growth rate [3] (p.422 in [3]).

The expression for the $\frac{dE}{dt}$ represented in the formula (79.1) in [3] cannot be directly derived from (1.1) and (1.2), as made for Eq. (2.5), but only is introduced on the basis of the relation (2.10) using the integral entropy balance equation (49.6) given in [3]. It is precisely the formula (2.5) for the quantity $\frac{dE}{dt}$ is a generalization of the formula (16.3) in [3] to case of divergent flows of the compressible medium, but not the formula (79.1), as stated in [3].

5. Thus, from (2.5) and (2.9) it follows that the negative definiteness of the dissipation rate of the integral kinetic energy and the corresponding positive definiteness of the growth rate of the integral entropy are possible for divergent flows of a compressible medium only under the condition of vanishing the second term on the right-hand sides of (2.5) and (2.9) when the relation (1.3) holds.

From Eq. (1.3) it follows that in (2.5) the decrease rate of the integral kinetic energy of divergent flows is now determined only by viscous dissipation, as for divergence-free flows (see (16.3) in [3]).

When Eq. (1.3) is fulfilled, the positively defined growth rate of the integral entropy in (2.9) turns out to be appreciably less than the growth rate of the integral entropy given by formula (49.6) in [3]. In fact, in (49.6) there is a term proportional to the second viscosity coefficient, while in (2.9) there is no such a term when the condition (1.3) is fulfilled. Providing the fulfillment of (1.3), this decrease in the entropy growth rate in (2.9) corresponds to the relative decrease in the kinetic energy dissipation rate in (2.5), as compared with the expression (79.1) in [3]. In this case there is a correspondence with the Prigozhin principle of minimum entropy production (see [9]).

Thus, for divergent flows of a compressible medium on the basis of the requirement of positive definiteness of the growth rate of the integral entropy in (2.9) and negative definiteness of the growth rate of the integral energy in (2.5) we obtain the additional equation (1.3) and this equation closes the system (1.1), (1.2). Therefore, for divergent flows of the compressible medium equation (1.3) must replace the divergence-free condition usually used to close the system (1.1), (1.2) in the case of the incompressible-medium approximation.

### 3. New solution to the NS equation

1. Relation (1.3) determines the exact mutual compensation of the normal pressure stresses and the normal viscous stresses of compressible divergent flow. As a result of this compensation, the second term on the right-hand side of Eq. (1.1) vanishes. In this case the equation (1.1) coincides exactly with the *n*-dimensional generalization of the Burgers equation (1.4).

The system (1.2), (1.4) is already closed and describes the evolution of the density and the velocity field of the medium in its inertial motion in the presence of damping related to only the action of shear viscous stresses corresponding to the nonzero right-hand side of Eq. (1.4).

At the same time, if in (1.4) the viscosity coefficient is equal to zero, then from (1.4) we can obtain the *n*-dimensional RH equation for which the exact vortical solution was obtained in [12, 13] (see also Appendix A). In what follows, we shall consider this solution and generalize it by taking the external friction and the effective volume viscosity into account on the base of equation (1.6).

We note that, as distinct from consideration of the vortical solutions carried out in the present study and [12-16], only the vortex-free solution to the 3-D Burgers equation was investigated earlier (when $\frac{\eta}{\rho} = \nu = const$ in (1.4)). This solution corresponds only to the potential flow and can be obtained using a modification of the nonlinear Cole-Hopf transform [34, 35].

With the aim to introduce effective volume viscosity (in addition to external friction in (1.6)) let the change $u_i \to u_i + V_i(t)$ be implemented in (1.6), where $V_i(t)$ is a random Gaussian delta-correlated-in-time velocity field for which the relations hold

$$\begin{aligned} &\langle V_i(t)V_j(\tau)\rangle = 2\nu\delta_{ij}\delta(t-\tau) \\ &\langle V_i(t)\rangle = 0 \end{aligned} \qquad (3.1)$$

In (3.1) $\delta_{ij}$ is the Kronecker delta, $\delta$ is Dirac-Heaviside delta-function, and the coefficient $\nu$ characterizes the action of the viscosity forces. In the general case the coefficient $\nu$ can be a function of time when describing the effective turbulent viscosity but also it can coincide with the constant kinematic viscosity coefficient when the random velocity field considered corresponds to molecular fluctuations. We will restrict our attention to consideration of the case of constant coefficient $\nu$ in (3.1).

As was noted in Introduction, this change in (1.6) related to introduction of a random velocity field corresponds to the method [18] of obtaining the stochastic NS equation not due to the use of a random force but by adding a random velocity to the velocity field which enters into the usual deterministic NS equation. As distinct from [18], in the present study we consider the situation in which such a random velocity field depends only on time (in [18] such a velocity field is called by the drift part of a large-scale random field) and taking this field into account is equivalent to introduction of a bulk viscosity force which coincides in the structure with the usual friction force in the NS equation.

From Eq. (1.6) averaged with allowance for (3.1) we can obtain the equation

$$\frac{\partial\langle u_i\rangle}{\partial t} + \left\langle u_j \frac{\partial u_i}{\partial x_j}\right\rangle = \nu\Delta\langle u_i\rangle - \mu\langle u_i\rangle \ , \qquad (3.2)$$

where the broken brackets denote the operation of averaging over the random Gaussian velocity field $V_i(t)$. In deriving Eq. (3.2) from (1.6), (3.1), in addition we use the following relation which is a consequence of the Furutsu - Novikov formula [36-38]:

$$\left\langle V_k \frac{\partial u_i}{\partial x_k} \right\rangle = -\nu\Delta\langle u_i \rangle \qquad (3.3)$$

Equation (3.2) for the case of zero external friction with $\mu = 0$ in (3.2) can also correspond to Eq. (1.4) if the equalities $\langle u_i \rangle = u_i$ and $\left\langle u_j \frac{\partial u_i}{\partial x_j} \right\rangle = u_j \frac{\partial u_i}{\partial x_j}$ are fulfilled (see [39]) and if in (1.4) we preliminarily perform the change $\frac{\eta}{\rho} \to \min(\frac{\eta}{\rho}) = \nu$. Such an uncoupling of correlations is possible in the case of exact separation of time scales related to large-scale inertial motions and motions with the characteristic scale of viscous dissipation [39, 26].

2. Rather than to solve approximately (e.g., see [39]) the closure problem to find the average velocity field $\langle u_i \rangle$ by considering directly Eq. (3.2), we shall use the initial equation from which just the equation (3.2) follows exactly. This initial equation (1.6) for the case $\mu = 0$ takes the form of the *n*-dimensional RH equation [9, 12]:

$$\frac{\partial u_i}{\partial t} + (u_j + V_j(t))\frac{\partial u_i}{\partial x_j} = 0 \qquad (3.4)$$

Indeed, if we take average in (3.4) on $\vec{V}(t)$ the equation (3.2) is exactly obtained if (3.3) is used.

As shown in Appendix A, in the case of an arbitrary dimensionality of the space (*n*=1, 2, 3, etc.) equation (3.4) has the following exact solution (see also [12-16]):

$$u_i(\vec{x},t) = \int d^n \xi u_{0i}(\vec{\xi})\delta(\vec{\xi} - \vec{x} + \vec{B}(t) + t\vec{u}_0(\vec{\xi}))\det \hat{A}, \qquad (3.5)$$

where $B_i(t) = \int_0^t dt_1 V_i(t_1)$, $\hat{A} \equiv A_{nm} = \delta_{nm} + t\frac{\partial u_{0n}}{\partial \xi_m}$, $\det \hat{A}$ - is the determinant of the matrix $\hat{A}$, and $u_{0i}(\vec{x})$ is an arbitrary smooth initial velocity field. The solution (3.5) satisfies Eq. (3.4) only at such times for which the determinant of the matrix $\hat{A}$ is positive for any values of the spatial coordinates, i.e., $\det \hat{A} > 0$. Therefore, everywhere in what follows we shall not use the symbol of absolute value in writing $\det \hat{A}$, if the opposite is not mentioned.

Only in the case of the potential initial velocity field the solution (3.5) is potential for all successive instants of time, corresponding to zero vortex field. On the contrary, in the case of nonzero initial vortex field the solution also determines the evolution of velocity with nonzero vortex field (see next section). In what follows, we shall restrict our attention to consideration of only the vortex solutions in (3.5). However, we note that in [27] it is precisely the potential solution to the two-dimensional RH equation (3.4) (when $\vec{B} = 0$ in (3.4)) was obtained in the Lagrangian representation which also exactly follows from (3.5) for *n*=2, as was already noted in Introduction in connection with the possibility of description of the solution of the Sivashinsky equation (I.2) using the potential solution in (3.5).

In the one-dimensional case (*n*=1) in (3.5) we have $\det \hat{A} = 1 + t\frac{du_{01}}{d\xi_1}$ and the solution (3.5) coincides exactly with the solutions obtained in [36, 37]. The solution (3.5) can be obtained if we use the integral representation for the implicit solution of Eq. (3.4) in the form $u_k(\vec{x},t) = u_{0k}(\vec{x} - \vec{B}(t) - t\vec{u}(\vec{x},t))$ with the use of the Dirac delta-function (see Appendix A or [12, 13]).

After averaging over the random field $B_i(t)$ (with the Gaussian probability density), from (3.5) we can obtain the exact solution of Eq. (3.2) (for $\mu = 0$) in the form:

$$\langle u_i \rangle = \int d^n \xi u_{0i}(\vec{\xi}) \left|\det \hat{A}\right| \frac{1}{(2\sqrt{\pi \nu t})^n} \exp\left[ -\frac{(\vec{x} - \vec{\xi} - t\vec{u}_0(\vec{\xi}))^2}{4\nu t} \right] \quad (3.6)$$

As distinct from (3.5), the average solution (3.6) of equation (3.2) is already arbitrary smooth on any unbounded time interval and not only providing the positiveness of the determinant of the matrix $\hat{A}$. The solution (3.6) is also the solution of NS equation (1.1) when the conditions (1.3) and (1.5) are take place in limit $\mu \to 0$.

3. If we neglect the viscosity forces when $\vec{B}(t) = 0$ in (3.5), the smooth solution (3.5) is defined, as was already noted, only under the condition $\det \hat{A} > 0$ [12, 13] (see Appendix A). This condition corresponds to a bounded time interval $0 \le t < t_0$, where the minimum limiting time $t_0$ of existence of the solution can be determined from the solution to the following *n*th-order algebraic equation (and successive minimization of the expression obtained, which depends on the spatial coordinates, with respect to these coordinates):

$$\begin{aligned}
&\det \hat{A}(t) = 1 + t\frac{du_{01}(x_1)}{dx_1} = 0, n = 1 \\
&\det \hat{A}(t) = 1 + t\,div\,\vec{u}_0 + t^2 \det \hat{U}_{012} = 0, n = 2 \\
&\det \hat{A}(t) = 1 + t\,div\,\vec{u}_0 + t^2(\det \hat{U}_{012} + \det \hat{U}_{013} + \det \hat{U}_{023}) + t^3 \det \hat{U}_0 = 0, n = 3
\end{aligned} \quad (3.7)$$

where $\det \hat{U}_0$ is the determinant of the 3×3 matrix $U_{0nm} = \frac{\partial u_{0n}}{\partial x_m}$, and

$\det \hat{U}_{012} = \frac{\partial u_{01}}{\partial x_1}\frac{\partial u_{02}}{\partial x_2} - \frac{\partial u_{01}}{\partial x_2}\frac{\partial u_{02}}{\partial x_1}$ is the determinant of a similar matrix in the two-dimensional case for the variables $(x_1, x_2)$. In this case $\det \hat{U}_{013}, \det \hat{U}_{023}$ are the determinants of the matrices in the two-dimensional case for the variables $(x_1, x_3)$ and $(x_2, x_3)$, respectively.

We note that in the two-dimensional case equation (3.7) exactly coincides with the collapse condition obtained in [27] in connection with the problem of propagation of a flame front investigated on the basis of the Sivashinsky equation (I.2). In this case for exact coincidence it is necessary to replace $t \to b(t) = \frac{U_s(\exp(\gamma_0 t) - 1)}{\gamma_0}$ in (3.7).

In the one-dimensional case, when *n*=1, from Eq. (3.7) we can obtain the minimum time of appearance of the singularity $t_0 = \frac{1}{\max\left|\frac{du_{01}(x_1)}{dx_1}\right|} > 0$. In particular, for the initial distribution

$u_{01}(x_1) = a\exp(-\frac{x_1^2}{L^2}), a > 0$ it follows that $t_0 = \frac{L}{a}\sqrt{\frac{e}{2}}$ obtained for the value $x_1 = x_{1\max} = \frac{L}{\sqrt{2}}$. In this case the singularity itself can be implemented only for positive values of the coordinate $x_1 > 0$ when equation (3.7) has a positive solution for time.

This means that singularity (collapse) of the smooth solution can never occur when the initial velocity field is nonzero only for negative values of the spatial coordinate $x_1 < 0$.

Similarly, we can also determine the vortex wave burst time $t_0$ for *n*>1. For (3.7) in the two-dimensional case (when the initial velocity field is divergence-free) for the initial stream function in the form $\psi_0(x_1, x_2) = a\sqrt{L_1 L_2}\exp(-\frac{x_1^2}{L_1^2} - \frac{x_2^2}{L_2^2}), a > 0$ we obtain that the minimum time of existence of the smooth solution is equal to $t_0 = \frac{e\sqrt{L_1 L_2}}{2a}$.

In the example considered this minimum time of existence of the smooth solution is implemented for the spatial variables corresponding to points on the ellipse $\frac{x_1^2}{L_1^2} + \frac{x_2^2}{L_2^2} = 1$.

In accordance with (3.7) the necessary condition of implementation of the singularity is the condition of existence of a real positive solution to a quadratic (when *n*=2) or cubic (when *n*=3) equation for the time variable *t*. For example, in the case of two-dimensional flow with the initial divergence-free velocity field $div\vec{u}_0 = 0$, in accordance with (3.7), the necessary and sufficient condition of implementation of the singularity (collapse) of the solution in finite time is the condition

$$\det U_{012} < 0 \quad (3.8)$$

For the example considered above from (3.8) there follows the inequality $\frac{x_1^2}{L_1^2} + \frac{x_1^2}{L_2^2} > \frac{1}{2}$. When this inequality is satisfied, for *n*=2 there exists a real positive solution to the quadratic equation in (3.7) for which the minimum collapse time $t_0 = \frac{e\sqrt{L_1 L_2}}{2a} > 0$ given above is obtained.

On the contrary, if the initial velocity field is defined in the form of a finite function which is nonzero only in the domain $\frac{x_1^2}{L_1^2} + \frac{x_2^2}{L_2^2} \le \frac{1}{2}$, then the inequality (3.8) is violated and the development of singularity in a finite time turns out already to be impossible and the solution remains smooth in unbounded time even regardless of the viscosity effects.

The condition of existence of a real positive solution of Eq. (3.7) (e.g., see (3.8)) is the necessary and sufficient condition of implementation of the singularity (collapse) of the solution, as distinct from the sufficient but not necessary integral criterion which was proposed in [25] (see formula (38) in [25]) and has the form:

$$(\frac{dI}{dt})_{t=0} = -\int d^3x div\vec{u}_0 \det{}^2\hat{U}_0 > 0; I = \int d^3x \det{}^2\hat{U} \qquad (3.9)$$

In fact, in accordance with this criterion proposed in [25], the collapse of the solution is not possible in the case of the initial divergence-free velocity field, i.e., when $div\vec{u}_0 = 0$. However, in this case the violation of criterion (3.9) does not exclude the possibility of the collapse of the solution by virtue of the fact that the criterion (3.9) does not determine the necessary condition of implementation of the collapse. Actually, in the example considered above (in determination of the minimum time of implementation of the collapse $t_0 = \frac{e\sqrt{L_1 L_2}}{2a}$) for two-dimensional compressible flow the initial condition corresponded just to the initial velocity field with $div\vec{u}_0 = 0$ in (3.7) when *n*=2.

4. On the basis of the solution (3.5), using (3.7) and the Lagrangian variables $\vec{a}$ (where $\vec{x} = \vec{x}(t,\vec{a}) = \vec{a} + t\vec{u}_0(\vec{a})$), we can represent the expression for the matrix of the first derivatives of the velocity $\hat{U}_{im} = \frac{\partial u_i}{\partial x_m}$ in the form:

$$\hat{U}_{im}(\vec{a},t) = \hat{U}_{0ik}(\vec{a}) A^{-1}_{km}(\vec{a},t) \tag{3.10}$$

In this case the expression (3.10) exactly coincides with the formula (30) given in [25] for the Lagrangian time evolution of the matrix of the first derivatives of the velocity which must satisfy the three-dimensional RH equation (3.4) (when $\vec{B}(t) = 0$ in (3.4)). In particular, in the one-dimensional case when *n*=1, in the Lagrangian representation from (3.5) and (3.7) we obtain a particular case of the formula (3.10):

$$\left(\frac{\partial u(x,t)}{\partial x}\right)_{x=x(a,t)} = \frac{\dfrac{du_0(a)}{da}}{1 + t\dfrac{du_0(a)}{da}}, \tag{3.11}$$

where $a$ is the coordinate of a fluid particle at the initial time $t = 0$.

The solution (3.11) also coincides with the formula (14) in [25] and describes the catastrophic process of collapse of a simple wave in a finite time $t_0$ whose estimate is given above on the basis of the solution to Eq. (3.7) in the case $n = 1$ with the use of the Euler variables.

## 4. Exact solution to the EH and RH equations

The velocity field (3.5) corresponds to the exact solution for the vortex field which has the following form in the two- and three-dimensional cases, respectively (see Appendix B or [12, 13]):

$$\omega(\vec{x},t) = \int d^2\xi\, \omega_0(\vec{\xi})\, \delta(\vec{\xi} - \vec{x} + \vec{B}(t) + t\vec{u}_0(\vec{\xi})) \tag{4.1}$$

$$\omega_i(\vec{x},t) = \int d^3\xi \left(\omega_{0i}(\vec{\xi}) + t\omega_{0j}\frac{\partial u_{0i}(\vec{\xi})}{\partial \xi_j}\right) \delta(\vec{\xi} - \vec{x} + \vec{B}(t) + t\vec{u}_0(\vec{\xi})) \tag{4.2}$$

where in (4.2) $\vec{\omega}_0 = rot\vec{u}_0$ and in (4.1) $\omega_0$ is the initial vorticity distribution in the two-dimensional case. The solution (4.2), (3.5) corresponds to the following exact expression for the helicity:

$$H = \omega_k u_k = \int d^3\xi \left(u_{0k}\omega_{0k} + t\omega_{0j}\frac{\partial}{\partial \xi_j}\left(\frac{\vec{u}_o^{\,2}}{2}\right)\right) \delta(\vec{\xi} - \vec{x} + \vec{B}(t) + t\vec{u}_0(\vec{\xi})) \tag{4.3}$$

The representations for the three-dimensional vortex and velocity fields (4.2) and (3.5) exactly satisfy (see Appendix B) the three-dimensional EH and RH equations (1.7) and (3.4) for the case of zero external friction with $\mu = 0$ in (1.7) and when in (1.7) we also make the change $\vec{u} \to \vec{u} + \vec{B}(t)$, as in (3.4).

After averaging carried out in (4.1) - (4.3) over the random Gaussian field $\vec{B}(t)$ and taking (3.1) into account, we obtain expressions in which it is necessary to substitute an exponential function with a normalizing factor, as that in (3.6), in place of the delta function in the integrands in (4.1) - (4.3). Only after this averaging the existence of not only the average vortex field and helicity themselves is ensured on any time interval, but also the existence of the corresponding higher derivatives and higher moments. In particular, this is valid for the enstrophy (integral of the vorticity square over the entire space) and the higher moments of the vortex field for which explicit analytic expressions will be elementary obtained in the next section without solving any closure problem.

2. In the Lagrangian variables, in the case $\vec{B}(t)=0$ the expressions corresponding to the Eulerian vortex and helicity fields (4.1), (4.2), and (4.3) can be represented in the form:

$$\omega(\vec{a},t)=\frac{\omega_0(\vec{a})}{\det\hat{A}(\vec{a},t)}, \quad (4.4)$$

$$\omega_i(\vec{a},t)=\frac{(\omega_{0i}(\vec{a})+t\omega_{0m}(\vec{a})\frac{\partial u_{0i}(\vec{a})}{\partial a_m})}{\det\hat{A}(\vec{a},t)}, \quad (4.5)$$

$$H(\vec{a},t)=\frac{(u_{0k}(\vec{a})\omega_{0k}(\vec{a})+t\omega_{0k}(\vec{a})\frac{\partial}{\partial a_{0k}}(\frac{\vec{u}_0^2(\vec{a})}{2}))}{\det\hat{A}(\vec{a},t)}, \quad (4.6)$$

where $\det\hat{A}=\det(\delta_{im}+t\frac{\partial u_{0i}(\vec{a})}{\partial a_m})$.

From (4.4) – (4.6) it follows that in the two- and three-dimensional cases for a Lagrangian fluid particle the singularity of both the vortex and the helicity takes place as $t\to t_0$ when $\det\hat{A}(\vec{a},t)\to 0$ and the value of finite time of existence of the corresponding smooth fields can be determined from the Lagrangian analog of the condition (3.11). In this case from (4.5) and (4.6) it follows that the three-dimensional effect of extension of the vortex lines leads only to a weaker power-law but not the explosive increase in the values of the vortex and helicity, as distinct from the catastrophic process of collapse of vortex waves in finite time $t_0$ just for the divergent flow of the compressible medium.

We note that in [26] (see the formula (23) in [26]) the representation of the solution to the EH equation (1.7) (when $\mu=0$ in (1.7)) was obtained in the form:

$$\omega_i=\frac{\omega_{0k}(\vec{a})}{J}\frac{\partial R_i(\vec{a},t)}{\partial a_k} \quad (4.7)$$

In (4.7) $J=\det\left(\frac{\partial x_n}{\partial a_m}\right)$ is the Jacobian of the transformation to the Lagrangian variables $\vec{a}$. In this case $\vec{\omega}_0(\vec{a})$ is a new Cauchy invariant (coinciding with the initial vorticity) characterized by zero divergence $\frac{\partial\omega_{0k}(\vec{a})}{\partial a_k}=0$, while $x_i=R_i(\vec{a},t)$ and $\frac{d\vec{R}}{dt}=\vec{V}_n(\vec{R},t)$, where $\vec{V}_n$ is the velocity component normal to the vorticity vector so that $div\vec{V}_n\neq 0$ for it [26].

As distinct from (4.4) and (4.5), the expression (4.7) does not give any explicit representation for the solution to the EH equation since in (4.7) no definite dependence is given for the Jacobian $J$ and the vector $\vec{R}$. At the same time, there is a structural correspondence between (4.7) and (4.4), (4.2). In the case of the inertial motion of Lagrangian fluid particles, for the Jacobian in (4.7) we can already use the explicit representation $J=\det\hat{A}$, where $\det\hat{A}$ can be determined from (3.7).

## 5. Enstrophy balance equation and the homogeneous friction

1. From (4.1), (4.2) we can obtain the exact closed description of the enstrophy of two- and three-dimensional flows of an ideal compressible medium in the form [12, 13] (see also Appendix B):

$$\Omega_2 \equiv \int d^2 x \omega^2(\vec{x},t) = \int d^2 \zeta \omega_0^2(\vec{\zeta}) / \det \hat{A} \qquad (5.1)$$

$$\Omega_3 \equiv \int d^3 x \omega_i^2(\vec{x},t) = \int d^3 \zeta (\omega_{0i} + t\omega_{0j} \frac{\partial u_{0i}}{\partial \zeta_j})^2 / \det \hat{A} \qquad (5.2)$$

In obtaining the expressions (5.1) and (5.2) there was no need for solving the closure problem usually existing in theory of turbulence. In the present study this problem can be got round owing to the relatively simple representation of the exact solution to the nonlinear EH equation for describing vortex flow.

The expressions (5.1) and (5.2) tend to infinity in finite time $t_0$, which can be determined from the solution of the algebraic equation (3.7) and the successive minimization of this solution in the spatial coordinates.

Using the exact solution to the EH equations in the form of (4.1) and (4.2), we can obtain the closed description of the time evolution of not only the enstrophy, as made in (5.1) and (5.2), but also for any higher moments of the vortex field.

For example, in the two-dimensional case, taking into account from (4.1) and (A.15) – (A.17) (see Appendix A) we obtain

$$\Omega_{2(m)} = \int d^2 x \omega^m = \int d^2 \xi \frac{\omega_0^m(\vec{\xi})}{\det^{m-1} \hat{A}}; \Omega_{2(2m)} = \int d^2 x \omega^{2m} = \int d^2 \xi \frac{\omega_0^{2m}(\vec{\xi})}{\det^{2m-1} \hat{A}}; m = 1,2,3,...$$

In Introduction we gave the estimate (I.3) for the relation between different moments of the three-dimensional vortex field which was obtained on the basis of the expressions of the same type from (4.2) and (A.15)- (A.17).

The estimate $\det \hat{A} \cong O(t_0 - t)$ implemented in the limit as $t \to t_0$ is also used in obtaining (I.3). In this case the minimum collapse time $t_0$ can be determined from (3.7).

2. We will now take into account the external friction. For this purpose it is necessary to consider the case with $\mu > 0$ in Eq. (1.7). In this case we can also obtain the exact solution from the expressions (3.5), (4.1), and (4.2) changing in them the time variable *t* by the variable $\tau = \frac{1 - \exp(-t\mu)}{\mu}$ (see (A.13) in Appendix A and [12, 13]). The new time variable $\tau$ now varies within the finite limits from $\tau = 0$ (when $t = 0$) to $\tau = \frac{1}{\mu}$ (as $t \to \infty$). This leads to the fact that in the case of fulfillment of the inequality

$$\mu > \frac{1}{t_0}, \qquad (5.3)$$

for given initial conditions the quantity $\det \hat{A} > 0$ in the denominator of the expressions (5.1) and (5.2) so that the denominator cannot vanish at any instant of time, since the necessary and sufficient condition of implementation of the singularity (3.7) will be not satisfied. The change $t \to \tau(t)$ must also be carried out in the condition (3.7).

Providing (5.3), the solution to the three-dimensional EH equation is smooth on an unbounded interval of time *t*. The corresponding analytic divergent vortical solution to the three-dimensional NS equation also remains smooth for any $t \geq 0$ if the condition (5.3) is satisfied. In NS equation it is necessary to take the relations (1.3) and (1.5) into account.

We note that under the formal coincidence of the parameters $\mu=-\gamma_0$ (see the Sivashinsky equation (I.2) in Introduction) the equality $\tau(t)=b(t)$ takes place providing the implementation of singularity (3.7) when *n*=2 and in accordance with the solution of the Sivashinsky equation in [27]).

3. For inviscid (ideal) incompressible fluid flows with the divergence-free velocity field the explosive growth of the enstrophy is characteristic of the three-dimensional flows only, while for the two-dimensional flows the enstrophy is an invariant. Another situation takes place for divergent flows of the compressible medium considered in the present study.

In fact, for divergent flows of an inviscid medium in the two- and three-dimensional cases we have the enstrophy balance equations which follow from the EH equation (see (B.1) in Appendix B, or when $\mu=0$ in (1.7)):

$$\frac{d\Omega_2}{dt}=-\int d^2\zeta\omega^2 div\bar{u}\,,\quad \frac{d\Omega_3}{dt}=2\int d^3\zeta\omega_i\omega_k\frac{\partial u_i}{\partial\zeta_k}-\int d^3\zeta\omega_i^2 div\vec{u} \qquad (5.4)$$

From (5.4) we can see that in the three-dimensional case the time evolution of the enstrophy $\Omega_3$ is determined by not only the effect of extension of the vortex filaments (the first term on the right-hand side) but also by the second term caused by the finiteness of divergence of the velocity field. For two-dimensional flow the time evolution of the enstrophy $\Omega_2$ is implemented only for nonzero divergence of the flow velocity field.

For the solution (3.5) the divergence of the velocity field takes the form [12, 13]:

$$\frac{\partial u_k}{\partial x_k}=\int d^n\zeta\frac{\partial\det\hat{A}}{\partial t}\delta(\vec{\zeta}-\vec{x}+\vec{B}(t)+t\vec{u}_0(\vec{\zeta})) \qquad (5.5)$$

The integral of the right-hand side of (5.5) over the entire unbounded space is equal to zero by virtue of fulfillment of identities (A.14) – (A.16) (see Appendix A) and the condition of vanishing the initial velocity field at infinity. As a result, the equality $\int d^n x div\vec{u}=0$ holds for the solution under consideration. This equality ensures fulfillment of the conservation law for the total fluid mass and the exact mutual integral compensation of the intensities of distributed sources and sinks.

For the three-dimensional case in (5.4), using (3.5), (A.14)- (A.16), (4.2), and (5.5), we can obtain exact expressions for the first and second terms on the right-hand side of (5.4) which describe the contribution of the effect of extension of the vortex filaments and of nonzero divergence of the velocity field, respectively, to the enstrophy growth rate. We can readily verify that the same expressions for the two terms mentioned-above can be also obtained by directly differentiating the expression for the enstrophy in (5.2). As a result, we obtain the equality

$$\frac{d\Omega_3}{dt}=2\int d^3\zeta(\omega_{0i}+t\omega_{0k}\frac{\partial u_{0i}}{\partial\zeta_k})\omega_{0m}\frac{\partial u_{0i}}{\partial\zeta_m}/\det\hat{A}-\int d^3\zeta(\omega_{0i}+t\omega_{0k}\frac{\partial u_{0i}}{\partial\zeta_k})^2\frac{\partial\det\hat{A}}{\partial t}/\det^2\hat{A} \qquad (5.6)$$

In (5.6) the first and second terms on the right-hand side correspond exactly to the first and second terms on the right-hand side of (5.4), respectively. From (5.6) it follows that in the inviscid case both terms tend to infinity as $t\to t_0$, when $\det\hat{A}\to 0$ in accordance with (3.7). The origin of both first terms on the right-hand sides of (5.6) and (5.4) relates to the effect of extension of the vortex filaments. Their integrands are proportional to $O(\frac{1}{\det\hat{A}})$. Evidently, this term in (5.6) makes a relatively lesser contribution to the explosive growth rate of the enstrophy as compared with the second term in (5.6) whose integrand is proportional to $O(\frac{1}{\det^2\hat{A}})$ and which exists only in the case of divergent flows with nonzero divergence of the flow velocity field.

Since, as noted above, taking the viscosity into account (in particular, in taking into account the external friction when the condition (5.3) is fulfilled) leads to regularization of even divergent solutions to the NS equation, we can expect that it is also possible for the solutions with nonzero divergence. For the latter a similar regularization will also appear possible by virtue of the relatively weaker (in the sense noted above) effect of extension of the vortex filaments as compared with the process of wave collapse in divergent flow. This question will be also discussed in the next section.

4. After averaging over a random field $B_k(t)$ with the Gaussian probability density, from (1.3) and (5.5) and taking (3.2) into account, we obtain the following representation for the pressure

$$\langle p\rangle = (\xi + \frac{\eta}{3})\int d^n\zeta \frac{\partial \det \hat{A}}{\partial t}\frac{1}{(2\sqrt{t\pi\nu})^n}\exp\left[-\frac{(\vec{\zeta}-\vec{x}+t\vec{u}_0(\vec{\zeta}))^2}{4\nu t}\right] \qquad (5.7)$$

The expression for the density corresponding to Eqs. (1.2) and solution (3.5) takes the form [12, 13]:

$$\rho = \int d^n\zeta\rho_0(\vec{\zeta})\delta(\vec{\zeta}-\vec{x}+\vec{B}(t)+t\vec{u}_0(\vec{\zeta})) \qquad (5.8)$$

After averaging in (5.8) and taking (3.2) into account, we obtain the following expression for the density of the medium which is smooth for any times:

$$\langle\rho\rangle = \int d^n\zeta\rho_0(\vec{\zeta})\frac{1}{(2\sqrt{t\pi\nu})^n}\exp\left[-\frac{(\vec{\zeta}-\vec{x}+t\vec{u}_0(\vec{\zeta}))^2}{4t\nu}\right] \qquad (5.9)$$

After replacing $\rho \to \omega, \rho_0 \to \omega_0$ in (5.9), we obtain an expression for the two-dimensional vortex field since the expressions (5.8) and (4.1) have the same structures.

### 6. To existence of the divergence-free solutions to the NS equation

The fact itself of the analytic representation of the smooth divergent solution to the NS equation (1.1) and equation (1.2) obtained above in the form (3.6), (5.7), (5.9) for an arbitrary smooth initial conditions, proves that the problem of existence and uniqueness is solved for this equation. It is of importance that in this case just the random Gaussian delta-correlated-in-time velocity field was introduced for simulating the viscosity effect. This leads to the effective viscosity force whose structure exactly corresponds to the structure of the viscosity force in the NS equation, as distinct from the higher derivatives than the Laplacian in the definition of the viscosity force considered in [19,20].

We will carry out a comparative analysis of the integral quantities for divergent and divergence-free flows which characterize the time evolution of the integral kinetic energy whose finiteness in [7] is the basic criterion of existence of the solution to the NS equation.

For this purpose, we will consider the integral kinetic energy balance equation (2.5) under the condition (1.3). In this case from (2.5) we can obtain the expression

$$\frac{dE}{dt} = -\eta F; \qquad F = \int d^3x(\frac{\partial u_i}{\partial x_k})^2 \qquad (6.1)$$

The form of the balance equation (6.1) exactly coincides with the form of the integral kinetic energy balance equation for divergence-free incompressible fluid flow given in [3] (see formula (16.3)). For divergent flow the functional $F$ in (6.1) is connected with the enstrophy $\Omega_3 = \int d^3x(rot\vec{u})^2$ by the following relation

$$F = \Omega_3 + D_3; D_3 = \int d^3 x (div\vec{u})^2 \qquad (6.2)$$

The right-hand side of (6.2) is deliberately greater than the value of the functional $F = F_0 = \Omega_3$ for the solution with zero divergence of the velocity field, when $D_3 = 0$ in (6.2).

For the exact solution obtained, the expression for the enstrophy $\Omega_3$ on the right-hand side of (6.2) has the form (5.2) and from (5.5) and (A.15) - (A.17) we obtained the following expression for the integral of the divergence square

$$D_3 = \int d^3\xi \left( \frac{\partial \det \hat{A}}{\partial t} \right)^2 / \det \hat{A} \qquad (6.3)$$

From a comparison of (6.3) and (5.2) it follows that in the neighborhood of the solution as $t \to t_0$ (see (3.7)) the values of the first and second terms on the right-hand side of (6.2) are of the same order of magnitude.

In addition, for the functional $F$ in (6.1) we can obtain the following upper estimate using the Cauchy-Schwartz-Bunyakovskii inequality:

$$F^2 = \left[ \int d^3 x \vec{u} \Delta \vec{u} \right]^2 \le \int d^3 x \vec{u}^2 \int d^3 x (\Delta \vec{u})^2 = \int d^3 x \vec{u}^2 \int d^3 x \left[ (rotrot\vec{u})^2 + (graddiv\vec{u})^2 \right] \quad (6.4)$$

In accordance with (6.2) - (6.4) the divergent flows have, all other factors being the same, the deliberately larger value of the functional $F$ as compared with the divergence-free flows for which there is no second term in the square brackets on the right-hand side of (6.4).

From our consideration the conclusion on existence of smooth divergence-free solutions to the NS equation may follow from the proved fact of existence of smooth divergent solutions to the NS equation on unbounded time interval when taking the effective viscosity (or the external friction into account providing (5.3))

## Discussion and Conclusions

Thus, in (3.6), (5.7), and (5.9) we represent the analytic solution to the NS equation (1.1) and the continuity equation (1.2) for divergent flows with nonzero divergence of the velocity field (5.5). In the three-dimensional case finiteness of the energy integral $E = \frac{1}{2} \int d^3\zeta \langle u \rangle^2$ follows explicitly from (3.6). This satisfies the main requirement in the formulation of the problem of existence of the solution to the NS equation [7]. Along with this, the requirement, stated in [7] for arbitrary smoothness of the solution on any time intervals when describing the velocity and pressure fields, is also fulfilled.

We note that for the exact solution (3.5) obtained the energy integral $E_{00} = \frac{1}{2} \int d^3 x \vec{u}^2 = \frac{1}{2} \int d^3 \xi \vec{u}_0^2 \det \hat{A} < \infty$ also remains finite without averaging (for example, in the case $\vec{B}(t) = 0$ ) for any finite instant of time, although in the limit as $t \to \infty$ the energy also tends to infinity in accordance with the power law $O(t^3)$ (see (3.7)). In this case the solution (3.5) can be continued for any finite time $t_* \ge t_0$ in the Sobolev space $H^0(R^3)$ . This means that in the case of an ideal medium the flow energy must satisfy the claim laid in [7].

However, in this case the enstrophy integral in (5.2) has already the explosive unbounded growth (in a finite time $t_0$, determined from (3.7)), when $\Omega_3 \cong O(\frac{1}{t_0 - t})$ in the case of a single real positive root of Eq. (3.7). This means that the obtained exact solution to the EH equation in the form (3.5) and (4.2) cannot already be continued in the Sobolev space $H^1(R^3)$ for time $t_* \geq t_0$ in the norm (I.1). Taking the viscosity into account makes it possible to avoid the singular behavior of the enstrophy and the higher moments of the vortex field. This means that the solutions to the EH and NS equations can be continued for any $t_* \geq t_0$ in the Sobolev space $H^q(R^3)$ already for any $q \geq 1$.

In formulation [7] of the problem of existence of the solution to the three-dimensional NS equation it was proposed to restrict consideration only to the case of solutions with the divergence-free velocity field. However, in [7] it was noted the importance of consideration of precisely three-dimensional flows for which the effect of extension of vortex filaments in finite time can lead to constraint of existence of the solution to the NS equation only in small.

The conclusion obtained concerning existence of smooth divergent solutions to the three-dimensional NS equation due to taking even a low viscosity into account indicates also the possibility of positive solution of the problem of existence of smooth divergence-free solutions on an unbounded time interval. In fact, as established in (5.6), the effect of extension of vortex filaments makes a quite smaller contribution to the implementation of singularity of the solution as compared with the collapse of the vortex wave in divergent compressible flow. Both inequality (6.4) and equality (6.2) for the variation rate of the integral kinetic energy also indicate this possibility.

We must note that the exact solution to the EH and RH equations obtained in [12-16] and here gives the closed description of the time evolution of the enstrophy and any other moments of the vortex, velocity, pressure, and density fields. The possibility of a closed statistical description of the regimes of turbulence without pressure (simulated by means of the nonlinear three-dimensional RH equation (3.4)) was first noted in 1991 in [12]. We note that the general theoretic-field approach to theory of turbulence without pressure was developed in 1995 in [42] by Polyakov for the three-dimensional RH equation with a random force of the type of white noise (delta-correlated-in-time), where the relation between the violation of Galilean invariance and intermittence was established. However, a particular solution of the closure problem was obtained only in the one-dimensional case in the form (see formula (41) in [42]) of the explicit expression for the probability distribution $w(u, y)$ of the velocity difference $u$ at points located at a distance $y$ from each other.

In the present study the approach which makes it possible to take exactly into account the pressure is developed. Owing to this, the analytic solution to the complete NS equation for flow of a viscous compressible medium is obtained. In this case the main problem of theory of turbulence [1] is actually solved. The solution obtained can give the exact representation for the mutual characteristic functional of the velocity and density fields of the medium (in this case the pressure field can be uniquely determined from (1.3)). Earlier, the solution of the main problem of theory of turbulence was considered to be unachievable for the compressible fluid and in [1] in this connection it was written: “Unfortunately, this general problem is too difficult and at present we cannot see even an approach to its complete solution” (see p. 177 in [1]).

Using the obtained exact solutions of the EH, RH, and NS equations the turbulent regimes can be also simulated on the basis of the randomization method for integrable hydrodynamic problems proposed by Novikov [43] and developed in [9]. For this purpose it is necessary to introduce the probability measure on an ensemble of implementation of the initial conditions which in this case must be considered as random functions.

The possibility of existence of the solution to the Navier-Stokes equation established in the present study is based on a new time-dependent analytic solution of this equation, earlier considered to be impossible [1, 7]. In this case it is revealed that for existence of the solution on an unbounded time interval it is necessary to take precisely the viscosity forces into account. On the other hand, the question of stability of the solution obtained must be considered on the basis of the existing results which testify that the destabilizing effects of viscosity leading to dissipative instability are also possible [44-47].

The authors wish to thank E.A. Novikov for his interest to study and fruitful advices, E.A. Kuznetsov for his attention to study, useful discussion, and information on studies [18, 25 - 27], V.V. Lebedev and A.G. Chefranov for the constructive critical comments and A.G. Kritsuk for his interest to our results.

## Appendix A
## Exact solution of n-D Riemann-Hopf (RH) equation (n=1, 2, 3)

### 1A.The procedure of the exact solution obtaining

The Riemann-Hopf (RH) equation in the n-dimensional space (n=1..3) is as follows:

$$\frac{\partial u_i}{\partial t}+u_l\frac{\partial u_i}{\partial x_l}=0 \qquad \text{(A.1)}$$

Equation (A.1) is obtained from the NS equation (1.1) if in (1.1) to neglect he right-hand side when the fluid particles move by inertia. (see examples of (A.1) application in the problems of astrophysics and hydrodynamics of granulated medium in [48-51]).

When the external friction coefficient tends to zero in the equation (1.6), $\mu\to 0$, the equation (1.6) also coincides with the RH equation (A.1).

In the unbounded space, the general Cauchy problem solution for the equation (A.1) under arbitrary smooth initial conditions $\vec{u}_0(\vec{x})$ satisfying (1.8), (1.9) may be obtained as follows (see also in [12, 13]).

The equation (A.1) may be represented in an implicit form as follows:

$$u_i(\vec{x},t)=u_{0i}(\vec{x}-t\vec{u}(\vec{x},t))=\int d^n\xi u_{0i}(\vec{\xi})\delta(\vec{\xi}-\vec{x}+t\vec{u}(\vec{x},t)) \qquad \text{(A.2)}$$

In (A.2), $\delta$ is the Dirac delta-function. Using known (see farther) properties of the delta-function, it is possible to express the delta-function in (A.2) with the help of an identity true for the very velocity field meeting the equation (A.1):

$$\delta(\vec{\xi}-\vec{x}+t\vec{u}(\vec{x},t))\equiv\delta(\vec{\xi}-\vec{x}+t\vec{u}_0(\vec{\xi}))\left|\det\hat{A}\right| \qquad \text{(A.3)}$$

In (A.3), the matrix $\hat{A}$ depends only on the initial velocity field and is as follows:

$$\hat{A}\equiv A_{km}=\delta_{km}+t\frac{\partial u_{0k}(\vec{\xi})}{\partial\xi_m} \qquad \text{(A.4)}$$

To infer (A.3), it is necessary to use the following delta-function property that is true for any smooth function $\vec{\Phi}(\vec{\xi})$:

$$\delta(\vec{\Phi}(\vec{\xi}))=\frac{\delta(\vec{\xi}-\vec{\xi}_0)}{\left|\det\left(\frac{\partial\Phi_k}{\partial\xi_m}\right)_{\vec{\xi}=\vec{\xi}_0}\right|} \qquad \text{(A.5)}$$

In (A.5), the values $\vec{\xi}_0$ are defined from the solution of the equation

$$\vec{\Phi}(\vec{\xi}_0)=0 \qquad \text{(A.6)}$$

To prove (A.5), it is necessary using Taylor series decomposition wrt $\vec{\xi}$ near $\vec{\xi}=\vec{\xi}_0$ for the argument of the delta-function $\vec{\Phi}(\vec{\xi})$ when in the limit $\vec{\xi}\to\vec{\xi}_0$ taking into account (A.6), we get

$$\delta(\Phi_k(\vec{\xi}_0)+(\frac{\partial\Phi_k}{\partial\xi_m})_{\vec{\xi}=\vec{\xi}_0}(\xi_m-\xi_{0m})+O(\vec{\xi}-\vec{\xi}_0)^2)=\delta((\frac{\partial\Phi_k}{\partial\xi_m})_{\vec{\xi}=\vec{\xi}_0}(\xi_m-\xi_{0m})) \quad (A.7)$$

Using variables substitution in the argument of the right-hand side of (A.7) (of the type $\hat{A}\vec{x}=\vec{y}$ and taking into account that $d\vec{x}=\frac{d\vec{y}}{|\det\hat{A}|}$ [52]), we get from the right-hand side of (A.7) the right-hand side of (A.5).

When in (A.5), $\vec{\Phi}(\vec{\xi})\equiv\vec{\xi}-\vec{x}+t\vec{u}_0(\vec{\xi})$ and $\det\frac{\partial\Phi_k}{\partial\xi_m}=\det A_{km}$ where $A_{km}$ is from (A.4) then the equation (A.6) is reduced to the following equation:

$$\vec{\xi}_0-\vec{x}+t\vec{u}_0(\vec{\xi}_0)=0 \quad (A.8)$$

Solution of the equation (A.8) follows:

$$\vec{\xi}_0=\vec{x}-t\vec{u}(\vec{x},t) \quad (A.9)$$

That can be verified substituting (A.9) into (A.8) and taking into account that the general implicit solution of the equation (A.1) can be represented as $\vec{u}(\vec{x},t)=\vec{u}_0(\vec{x}-t\vec{u}(\vec{x},t))$ that is used in (A.2).

Let us use a known property of the delta-function that for any smooth function $\vec{f}(\vec{x})$, the following equality $\vec{f}(\vec{x})\delta(\vec{x}-\vec{x}_0)=\vec{f}(\vec{x}_0)\delta(\vec{x}-\vec{x}_0)$ holds. That is why, in the general case, it is possible multiplying the both sides of (A.5) by $\left|\det\frac{\partial\Phi_k(\vec{\xi})}{\partial\xi_m}\right|$ getting the following:

$$\delta(\vec{\xi}-\vec{\xi}_0)=\delta(\vec{\Phi}(\vec{\xi}))\left|\det\frac{\partial\Phi_k(\vec{\xi})}{\partial\xi_m}\right| \quad (A.10)$$

From (A.10) and (A.9), identical holding of the equality (A.3) follows.

Taking into account (A.3), from (A.2), we get an exact general (for any smooth initial velocity fields) solution of the Cauchy problem for the equation (A.1) as

$$u_i(\vec{x},t)=\int d^n\xi u_{0i}(\vec{\xi})\delta(\vec{\xi}-\vec{x}+t\vec{u}_0(\vec{\xi}))\det\hat{A}, \quad (A.11)$$

where $\det\hat{A}=\det(\delta_{mk}+t\frac{\partial u_{0m}(\vec{\xi})}{\partial\xi_k})$. That solution of the equation (A.1) is considered under the following condition:

$$\det\hat{A}>0. \quad (A.12)$$

That is why, sign of $\det\hat{A}$ is absent in (A.11). The condition (A.12) provides smoothness of the solution only on the finite time interval defined above from (3.7).

We can check that the very (A.11) under condition (A.12) exactly satisfies the equation (A.1) (see farther item 2) by direct substitution of (A.11) in (A.1). For example, in the one-dimensional case (n=1), the solution of the equation (A.1) in the form of (A.11) when $\det\hat{A}=1+t\frac{\partial u_{01}}{\partial\xi_1}$, was earlier obtained in [40, 41]. Solution (A.11) describes not only potential but also vortex solutions of the equation (A.1) in two- and three-dimensional cases for any smooth initial velocity field $\vec{u}_0(\vec{x})$ that was not known earlier for the solutions of the equation (A.1) [48-51].

Solution (A.11) of the equation (A.1) allows getting an exact solution of the equation (3.4) if in (A.11) to make a substitution: $\vec{x}\to\vec{x}-\vec{B}(t)$ that yields equation (3.4) representation as in (3.5).

The solution (A.11) also can describe an exact solution of the equation (1.6) for $\mu > 0$ if in (A.11) to substitute

$$t \to \frac{1-\exp(-t\mu)}{\mu} \qquad \text{(A.13)}$$

**2A.The direct validation of the solution**

To verify the solution (A.11) satisfies the equation (A.1), let us substitute (A.11) in the equation (A.1). Then we get from (A.1):

$$\int d^n\xi \left[ u_{0i}(\vec{\xi})\frac{\partial \det \hat{A}}{\partial t}\delta(\vec{\xi}-\vec{x}+t\vec{u}_0(\vec{\xi})) - u_{0i}u_{0m}\det \hat{A}\frac{\partial \delta(\vec{\xi}-\vec{x}+t\vec{u}_0(\vec{\xi}))}{\partial x_m}\right] + \int d^n\xi \int d^n\xi_1 F = 0, \quad \text{(A.14)}$$

where $F \equiv u_{0m}(\vec{\xi}_1)\det \hat{A}(\vec{\xi}_1)\delta(\vec{\xi}_1-\vec{x}+t\vec{u}_0(\vec{\xi}_1))u_{0i}(\vec{\xi})\det \hat{A}(\vec{\xi})\dfrac{\partial \delta(\vec{\xi}-\vec{x}+t\vec{u}_0(\vec{\xi}))}{\partial x_m}$.

To transform sub-integral expression in (A.14), the following identities shall be used:

$$\frac{\partial \delta(\vec{\xi}-\vec{x}+t\vec{u}_0(\vec{\xi}))}{\partial x_m} = -A^{-1}_{km}\frac{\partial \delta(\vec{\xi}-\vec{x}+t\vec{u}_0(\vec{\xi}))}{\partial \xi_k}, \qquad \text{(A.15)}$$

$$\frac{\partial \det \hat{A}}{\partial t} \equiv \frac{\partial u_{0m}}{\partial \xi_k}A^{-1}_{km}\det \hat{A}, \qquad \text{(A.16)}$$

$$\frac{\partial}{\partial \xi_k}(A^{-1}_{km}\det \hat{A}) \equiv 0 \qquad \text{(A.17)}$$

The identity (A.15) is obtained from the relationship (obtained by differentiating the delta-function having argument as a given function of $\vec{\xi}$ ) $\dfrac{\partial \delta(\vec{\xi}-\vec{x}+t\vec{u}_0(\vec{\xi}))}{\partial \xi_k} = -\dfrac{\partial \delta(\vec{\xi}-\vec{x}+t\vec{u}_0(\vec{\xi}))}{\partial x_l}A_{lk}$ after multiplying its both sides by the inverse matrix $A^{-1}_{km}$ (where $A_{lk}A^{-1}_{km} = \delta_{lm}$ и $\delta_{lm}$ is the unity matrix, or the Kronecker delta).

Validity of the identities (A.16) and (A.17) is proved by the direct checking. In the 1-dimensional case, when $\hat{A} = 1 + t\dfrac{du_{01}}{d\xi_1} = \det \hat{A}; \hat{A}^{-1} = (\det \hat{A})^{-1}$, it obviously follows directly from (A.16), (A.17).

Farther, in item 3, the proof of the identities (A.16) and (A.17) the two- and three-dimensional cases is given.

Taking into account (A.15)- (A.17), from (A.14), we get

$$\int d^n\xi \delta(\vec{\xi}-\vec{x}+t\vec{u}_0(\vec{\xi}))A^{-1}_{km}\det \hat{A}(u_{0i}\frac{\partial u_{0m}}{\partial \xi_k} - \frac{\partial}{\partial \xi_k}(u_{0i}u_{0m})) + \int d^n\xi \int d^n\xi_1 F_1 = 0 \qquad \text{(A.18)}$$

Where the sub-integral expression in the second term of the left-hand side of (A.18) is as follows:

$$F_1 = u_{0m}(\vec{\xi}_1)\frac{\partial u_{0i}(\vec{\xi})}{\partial \xi_k}\det \hat{A}(\vec{\xi}_1)\det \hat{A}(\vec{\xi})A^{-1}_{km}(\vec{\xi})\delta(\vec{\xi}-\vec{x}+t\vec{u}_0(\vec{\xi}))\delta(\vec{\xi}_1-\vec{x}+t\vec{u}_0(\vec{\xi}_1)) \quad \text{(A.19)}$$

To transform (A.19), it is necessary using the following identities:

$$\delta(\vec{\xi}-\vec{x}+t\vec{u}_0(\vec{\xi}))\delta(\vec{\xi}_1-\vec{x}+t\vec{u}_0(\vec{\xi}_1)) \equiv \delta(\vec{\xi}-\vec{x}+t\vec{u}_0(\vec{\xi}))\delta(\vec{\xi}_1-\vec{\xi}+t(\vec{u}_0(\vec{\xi}_1)-\vec{u}_0(\vec{\xi}))) \quad \text{(A.20)}$$

$$\delta(\vec{\xi}_1-\vec{\xi}+t(\vec{u}_0(\vec{\xi}_1)-\vec{u}_0(\vec{\xi}))) \equiv \frac{\delta(\vec{\xi}_1-\vec{\xi})}{\det \hat{A}} \qquad \text{(A.21)}$$

In (A.21), as it is noted above, $\det \hat{A} > 0$, and that is why, the sign is not used in the denominator of (A.21).

The identity (A.20) is a consequence of the noted above property of the delta-function (see discussion before the formula (A.10)).
To infer the identity (A.21), it is necessary considering in the argument of the delta-function a Taylor series decomposition of the function

$u_{0k}(\vec{\xi}_1)=u_{0k}(\vec{\xi})+(\frac{\partial u_{0k}(\vec{\xi}_1)}{\partial \xi_{1m}})_{\vec{\xi}_1=\vec{\xi}}(\xi_{1m}-\xi_m)+O(\vec{\xi}_1-\vec{\xi})^2$ near the point $\vec{\xi}_1=\vec{\xi}$ . Then the left-hand side of (A.21) has the form $\delta(\hat{A}(\vec{\xi}_1-\vec{\xi}))$ similar to that of the right-hand side of (A.7) and according to (A.5), we get from here the identity (A.21).
After application of the identity (A.21) to the expression (A.19) defining the form of the second term in (A.18), from (A.18), we get:

$$\int d^n\xi\delta(\vec{\xi}-\vec{x}+t\vec{u}_0(\vec{\xi}))A^{-1}_{km}\det\hat{A}\left[u_{0i}\frac{\partial u_{0m}}{\partial \xi_k}-\frac{\partial}{\partial \xi_k}(u_{0i}u_{0m})+u_{0m}\frac{\partial u_{0i}}{\partial \xi_k}\right]=0 \qquad (A.22)$$

Equality (A.22) holds identically due to the identical equality to zero of the expression in the brackets in the sub-integral expression in (A.22).
Thus, we have proved that (A.11) exactly satisfies the RH equation
(A.1) for any smooth initial velocity fields on the finite time interval under condition $\det\hat{A}>0$ discussed in (3.7).

**3A.The validation of identities (A.16), (A.17)**

In the 2-dimensional case, the elements of the inverse matrix $A^{-1}_{km}$ and the determinant of the matrix $\hat{A}$ are:

$$A^{-1}_{11}=\frac{1+t\partial u_{02}/\partial\xi_2}{\det\hat{A}};A^{-1}_{12}=-\frac{t\partial u_{01}/\partial\xi_2}{\det\hat{A}};A^{-1}_{21}=-\frac{t\partial u_{02}/\partial\xi_1}{\det\hat{A}};A^{-1}_{22}=\frac{1+t\partial u_{01}/\partial\xi_1}{\det\hat{A}} \qquad (A.23)$$

$$\det\hat{A}=1+t(\frac{\partial u_{01}}{\partial\xi_1}+\frac{\partial u_{02}}{\partial\xi_2})+t^2(\frac{\partial u_{01}}{\partial\xi_1}\frac{\partial u_{02}}{\partial\xi_2}-\frac{\partial u_{01}}{\partial\xi_2}\frac{\partial u_{02}}{\partial\xi_1}) \qquad (A.24)$$

Here, (A.24) corresponds to the formula (3.7) for n=2.
Using (A.23), it is possible to show that the following equality holds (in the left-hand side of (A.25), summation is assumed on the repeating indices from 1 to 2)

$$\frac{\partial u_{0m}}{\partial\xi_k}A^{-1}_{km}\det\hat{A}=\frac{\partial u_{01}}{\partial\xi_1}+\frac{\partial u_{02}}{\partial\xi_2}+2t(\frac{\partial u_{01}}{\partial\xi_1}\frac{\partial u_{02}}{\partial\xi_2}-\frac{\partial u_{01}}{\partial\xi_2}\frac{\partial u_{02}}{\partial\xi_1}) \qquad (A.25)$$

From (A.24), it follows that the right-hand side of (A.25) exactly matches $\frac{\partial\det\hat{A}}{\partial t}$ obtained when differentiating over time in (A.24). This proves the identity of (A.16) in the 2-dimensional case.
To prove the identity (A.17), let us introduce

$$B_m=\frac{\partial}{\partial\xi_k}(A^{-1}_{km}\det\hat{A}) \qquad (A.26)$$

Using (A.23), one gets from (A.26)

$$B_1=\frac{\partial}{\partial\xi_1}(1+t\frac{\partial u_{02}}{\partial\xi_2})-\frac{\partial}{\partial\xi_2}(t\frac{\partial u_{02}}{\partial\xi_1})\equiv 0 \qquad (A.27)$$

$$B_2=\frac{\partial}{\partial\xi_1}(-t\frac{\partial u_{01}}{\partial\xi_2})+\frac{\partial}{\partial\xi_2}(1+t\frac{\partial u_{01}}{\partial\xi_1})\equiv 0 \qquad (A.28)$$

The identities (A.27), (A.28) confirm the truth of the identity (A.17) in the 2-dimensional case.

Similarly, the identity (A.17) is proved in the 3-dimensional case. For that, we need the following representation of the entries of the inverse matrix $\hat{A}^{-1}$ [53]:

$$A_{11}^{-1}=\frac{1}{\det\hat{A}}\left[(1+t\frac{\partial u_{02}}{\partial\xi_2})(1+t\frac{\partial u_{03}}{\partial\xi_3})-t^2\frac{\partial u_{02}}{\partial\xi_3}\frac{\partial u_{03}}{\partial\xi_2}\right];\ A_{12}^{-1}=\frac{1}{\det\hat{A}}\left[t^2\frac{\partial u_{01}}{\partial\xi_3}\frac{\partial u_{03}}{\partial\xi_2}-t(1+t\frac{\partial u_{03}}{\partial\xi_3})\frac{\partial u_{01}}{\partial\xi_2}\right];$$

$$A_{13}^{-1}=\frac{1}{\det\hat{A}}\left[t^2\frac{\partial u_{01}}{\partial\xi_2}\frac{\partial u_{02}}{\partial\xi_3}-t(1+t\frac{\partial u_{02}}{\partial\xi_2})\frac{\partial u_{01}}{\partial\xi_3}\right];A_{21}^{-1}=\frac{1}{\det\hat{A}}\left[t^2\frac{\partial u_{02}}{\partial\xi_3}\frac{\partial u_{03}}{\partial\xi_1}-t(1+t\frac{\partial u_{03}}{\partial\xi_3})\frac{\partial u_{02}}{\partial\xi_1}\right];$$

$$A_{22}^{-1}=\frac{1}{\det\hat{A}}\left[(1+t\frac{\partial u_{01}}{\partial\xi_1})(1+t\frac{\partial u_{03}}{\partial\xi_3})-t^2\frac{\partial u_{01}}{\partial\xi_3}\frac{\partial u_{03}}{\partial\xi_1}\right];A_{23}^{-1}=\frac{1}{\det\hat{A}}\left[t^2\frac{\partial u_{01}}{\partial\xi_3}\frac{\partial u_{02}}{\partial\xi_1}-t(1+t\frac{\partial u_{01}}{\partial\xi_1})\frac{\partial u_{02}}{\partial\xi_3}\right];$$

$$A_{31}^{-1}=\frac{1}{\det\hat{A}}\left[t^2\frac{\partial u_{02}}{\partial\xi_1}\frac{\partial u_{03}}{\partial\xi_2}-t(1+t\frac{\partial u_{02}}{\partial\xi_2})\frac{\partial u_{03}}{\partial\xi_1}\right];A_{32}^{-1}=\frac{1}{\det\hat{A}}\left[t^2\frac{\partial u_{01}}{\partial\xi_2}\frac{\partial u_{03}}{\partial\xi_1}-t(1+t\frac{\partial u_{01}}{\partial\xi_1})\frac{\partial u_{03}}{\partial\xi_2}\right]$$

$$A_{33}^{-1}=\frac{1}{\det\hat{A}}\left[(1+t\frac{\partial u_{01}}{\partial\xi_1})(1+t\frac{\partial u_{02}}{\partial\xi_2})-t^2\frac{\partial u_{01}}{\partial\xi_2}\frac{\partial u_{02}}{\partial\xi_1}\right] \qquad \text{(A.29)}$$

From (A.26), in the 3-dimensional case, we get on the base of (A.29) that all three components of the vector $B_m\equiv 0$. For each $m=1,2,3$, we get identical zeroing separately for the sum of terms proportional to $t$ and separately for the sum of the terms proportional to $t^2$.

For example, the sum if the terms proportional to $t$ in the expression for $B_1$ is as follows $t\left[\frac{\partial}{\partial\xi_1}(\frac{\partial u_{02}}{\partial\xi_2}+\frac{\partial u_{03}}{\partial\xi_3})-\frac{\partial^2 u_{02}}{\partial\xi_2\partial\xi_1}-\frac{\partial^2 u_{03}}{\partial\xi_3\partial\xi_1}\right]\equiv 0$, and similar for the sum of the twelve terms proportional to $t^2$. Thus, the identity (A.17) is also proved in the 3-dimensional case.

Proof of the identity (A.16) also is possible in 3-D case on the base of (A.29) and (3.7) but is related to the cumbersome transformations.

## Appendix B
## The exact solution of 3-D EH equation

### 1B.The procedure of the solution obtaining

From the RH equation (A.1), after applying to it of the rotor operation, we get the EG equation for the vortex field in the 3-dimensional case is as follows:

$$\frac{\partial\omega_i}{\partial t}+u_k\frac{\partial\omega_i}{\partial x_k}=\omega_k\frac{\partial u_i}{\partial x_k}-\omega_i\frac{\partial u_k}{\partial x_k} \qquad \text{(B.1)}$$

In (B.1), $\omega_i=\varepsilon_{ijl}\frac{\partial u_l}{\partial x_j}$ is the vortex field, $\varepsilon_{ijl}$ is the absolute (perfect) anti-symmetric unity tensor of rank three.

Modification of the EG equation (1.7) in the case of zero external friction $\mu=0$ coincides with the equation (B.1).

Let us show that an exact Cauchy problem solution for the equation (B.1) for arbitrary smooth initial velocity $\vec{u}_0(\vec{x})$ and vortex $\vec{\omega}_0(\vec{x})=rot\vec{u}_0$ fields is the following expression

$$\omega_i(\vec{x},t)=\int d^3\xi(\omega_{0i}(\vec{\xi})+t\omega_{0k}\frac{\partial u_{0i}}{\partial\xi_k})\delta(\vec{\xi}-\vec{x}+t\vec{u}_0(\vec{\xi})), \qquad \text{(B.2)}$$

that is obtained by applying rotor operator to the velocity field (A.11) being an exact solution of the 3-dimensional RH equation (A.1).

Let us prove that (B.2) is derived from (A.11) in the 3-dimensional case after applying rotor operator to (A.11). Taking into account (A.15) and (A.17), we get:

$$\omega_i(\vec{x},t) = \varepsilon_{iml} \frac{\partial u_l(\vec{x},t)}{\partial x_m} = \int d^3\xi \delta(\vec{\xi} - \vec{x} + t\vec{u}_0(\vec{\xi})) D_i \,, \qquad \text{(B.3)}$$

The vector $D_i$ in (B.3) is defined as followse:

$$D_i = \varepsilon_{iml} \frac{\partial u_{0l}(\vec{\xi})}{\partial \xi_k} A^{-1}_{km} \det \hat{A} \qquad \text{(B.4)}$$

On the base of the matrix entries (A.29) of the inverse matrix $\hat{A}^{-1}$ in the 3-dimensional case, let us prove the following identity for the vector $D_i$ from (B.4):

$$D_i \equiv \omega_{0i}(\vec{\xi}) + t\omega_{0k}(\vec{\xi}) \frac{\partial u_{0i}(\vec{\xi})}{\partial \xi_k} \qquad \text{(B.5)}$$

For example, for $D_1$ (if $i = 1$ in (B.4)), the right-hand side of (B.4) is

$$\varepsilon_{1ml} \frac{\partial u_{0l}}{\partial \xi_k} A^{-1}_{km} \det \hat{A} = \frac{\partial u_{03}}{\partial \xi_2} - \frac{\partial u_{02}}{\partial \xi_3} + t\left[ \frac{\partial u_{01}}{\partial \xi_1} \left( \frac{\partial u_{03}}{\partial \xi_2} - \frac{\partial u_{02}}{\partial \xi_3} \right) + \frac{\partial u_{02}}{\partial \xi_1} \frac{\partial u_{01}}{\partial \xi_3} - \frac{\partial u_{01}}{\partial \xi_2} \frac{\partial u_{03}}{\partial \xi_1} \right] \qquad \text{(B.6)}$$

And the right-hand side of (B.6) identically equals to the right-hand side of (B.5) when $i = 1$ in (B.5).
Similarly, it is proved right-hand sides of (B.4) and (B.5) identical equality for $i = 2$ and 3.
After substitution of (B.5) into (B.3), we get an exact vortex solution of the Cauchy problem for the EG equation (B.1) in the form of (B.2).

In the case of modification of the equation (B.1) accounting for the effects of the volumetric vorticity related with the substitution $\vec{u} \to \vec{u} + \vec{V}(t)$ in (B.1), respective solution of the modified equation is obtained by the substitution in (A.11) and (B.2) $\vec{x} \to \vec{x} - \vec{B}(t), \vec{B}(t) = \int_0^t dt_1 \vec{V}(t_1)$ and has a form of (3.5) and (4.2), respectively.
An exact Cauchy problem solution for the 3-dimensional EG equation modification accounting for the external friction (1.7) is obtained from (B.2) under substitution in (B.2) of the form (A.13). Thus the obtained solution under condition (5.3) remains smooth for any time instance t>0.

**2B.The validation of the solution**

Let us show that (B.2) is an exact solution of the 3-dimensional EG equation (B.1) on the base of the direct substitution of (B.2) and (A.11) into (B.1). After that substitution, accounting for the identities (A.15), (A.17) and (A.21), we get (omitting transformations similar to (A.14), (A.18) and (A.22))

$$\int d^3\xi \delta(\vec{\xi} - \vec{x} + t\vec{u}_0(\vec{\xi})) \left[ \frac{\partial u_{0i}}{\partial \xi_k} \left( \omega_{0k} - A^{-1}_{km} \left( \omega_{0m} + t\omega_{0j} \frac{\partial u_{0m}}{\partial \xi_j} \right) \right) \right] = 0 \qquad \text{(B.7)}$$

Truth of (B.7) follows from the identity

$$\omega_{0k} \equiv A^{-1}_{km} \left( \omega_{0m} + t\omega_{0j} \frac{\partial u_{0m}}{\partial \xi_j} \right) \qquad \text{(B.8)}$$

The identity (B.8) is proved by multiplication of the both sides of (B.8) by the matrix $A_{ik}$. Taking into account definition of the inverse matrix, $A_{ik}A_{km}^{-1}=\delta_{im}$ , we get $A_{ik}\omega_{0k}=\omega_{0i}+t\omega_{0j}\frac{\partial u_{0i}}{\partial \xi_j}$. Bu by the definition of the matrix, $A_{ik}=\delta_{ik}+t\frac{\partial u_{0i}}{\partial \xi_k}$ , meanwhile, actually holds the following equality

$$A_{ik}\omega_{0k}=\omega_{0i}+t\omega_{0k}\frac{\partial u_{0i}}{\partial \xi_k}.$$

**3B.Conclusion**

Thus, it is proved that the vortex field in the form (B.2) and respective velocity field (A.11) are the exact solution of the Cauchy problem for the 3-dimensional EG equation (B.1). That solution preserves smoothness only in the final interval of time defined from (3.7). However, if introducing even the external friction under condition (5.3), corresponding modification of the exact solution defined by the substitution of (A.13) into (B.2) and (A.11), already is found arbitrary smooth for any time instance unbounded by the value. Similar conclusion is true for the mean vortex and velocity fields when arbitrary small effective viscosity is introduced (that is defined by the random Gaussian delta-correlated velocity field).

The exact solution obtained for the equations (A.1) and (B.1) as (A.11) and (B.2), for example, allows application in the astro-physics problems [29, 48], and also for hydrodynamics of the granulated media [49-51]. Actually, the vortical solutions of the RH equation (A.1) can more relevant describe the process in the system that the potential solutions of the equation [48] (see [50]). Also, it is interesting to conduct comparison of the results obtained on the base of the exact solutions (A.11) and (B.2) versus the results of theoretical [54] and numerical studies [29, 55] of the correlation characteristics of the compressible turbulence.